\documentclass[10pt,authoryear]{article}
\usepackage{appendix}
\usepackage{setspace}
\usepackage[font=small,  margin=2cm]{caption}
\usepackage[tbtags]{amsmath}
\usepackage{amsthm,amssymb,amsfonts}
\usepackage[numbers]{natbib}
\usepackage{multirow}
\usepackage{lscape}
\usepackage{subfigure}
\usepackage{makecell}
\usepackage{exscale}
\usepackage{booktabs}
\usepackage{array}
\usepackage{fullpage}
\usepackage{url}
\usepackage{algorithm}
\usepackage{algpseudocode}
\usepackage{bm}
\usepackage{smile}
\usepackage{mathtools}
\usepackage{wrapfig}
\usepackage{lipsum}
\usepackage{mathrsfs}
\usepackage{relsize}
\usepackage{dsfont}
\usepackage{multirow}
\usepackage{apalike}
\usepackage{chngcntr}
\usepackage[usenames,dvipsnames,svgnames,table]{xcolor}
\usepackage[colorlinks=true,
linkcolor=blue,
urlcolor=blue,
citecolor=blue]{hyperref}
\usepackage{ulem}

\newcommand{\bargmax}{\mathop{\mathrm{arg\ max}}}

\numberwithin{equation}{section}
\numberwithin{theorem}{section}
\numberwithin{corollary}{section}
\counterwithout{asmp}{section}
\numberwithin{definition}{section}

\begin{document}

\title{\LARGE  Simultaneous  Cluster Structure Learning and
Estimation of Heterogeneous Graphs for Matrix-variate fMRI Data}

	\author{Dong Liu\thanks{ Shanghai University of Finance and Economics, Shanghai, China; },~~Changwei Zhao\thanks{Zhongtai Institute for Financial Studies, Shandong University, Jinan, China; Corresponding author, Email:{\tt heyong@sdu.edu.cn}.},~~Yong He,\footnotemark[2]~~ Lei Liu\thanks{Division of Biostatistics, Washington University in St.Louis, St. Louis, USA;},~~Ying Guo\thanks{Department of Biostatistics and Bioinformatics, Rollins School of Public Health, Emory University, Atlanta, USA;},~~Xinsheng Zhang\thanks{Department of Statistics, School of Management, Fudan University, Shanghai,  China; }}	
	\date{}	
	\maketitle

\textbf{Abstract:}  Graphical models play an important role in neuroscience studies, particularly in brain connectivity analysis.
Typically, observations/samples are from several heterogenous groups and the group membership of each observation/sample is unavailable, which poses a great challenge for graph structure learning. In this article, we propose a method which can achieve Simultaneous Clustering and Estimation of Heterogeneous Graphs (briefly denoted as SCEHG) for matrix-variate function Magnetic Resonance Imaging (fMRI) data. Unlike the conventional clustering methods which rely on the mean differences of various groups, the proposed SCEHG method fully exploits the group differences of  conditional   dependence relationships among brain regions for learning cluster structure. In essence, by constructing individual-level between-region network measures, we formulate clustering as penalized regression with grouping and sparsity pursuit, which transforms the unsupervised learning into supervised learning. An ADMM algorithm is proposed to solve the corresponding optimization problem. We also propose a generalized criterion to specify the number of clusters. Extensive simulation studies illustrate the superiority of the SCEHG method over some state-of-the-art methods in terms of both clustering  and graph recovery accuracy. We also apply the SCEHG procedure to analyze fMRI data associated with ADHD (abbreviated for Attention Deficit Hyperactivity Disorder), which illustrate its empirical usefulness. An R package ``SCEHG" to implement the method is available at \url{https://github.com/heyongstat/SCEHG.}

\vspace{0.2em}

	
\section{Introduction}
In neuroscience studies, Brain Connectivity Analysis (BCA) has been at the foreground to uncover the potential pathogenic mechanism of mental disease such as Alzheimer's disease. Graphical models, which capture the (conditional) dependence relationships among a set of variables,  particularly suit to  BCA  and serves as a powerful tool in neuroscience studies. One of the most popular graphical models is the Gaussian Graphical Model (GGM), and it is well known that recovering the structure of a GGM is equivalent to finding the support of the corresponding precision matrix. In the high-dimensional case where the number of covariates can be much larger than the sample size, various penalized methods have been proposed to estimate the GGM, see for example, \cite{meinshausen2006high,yuan2007model,cai2011constrained,He2016High}.

\begin{figure}[!h]
	\centerline{\includegraphics[width=20cm,height=8cm]{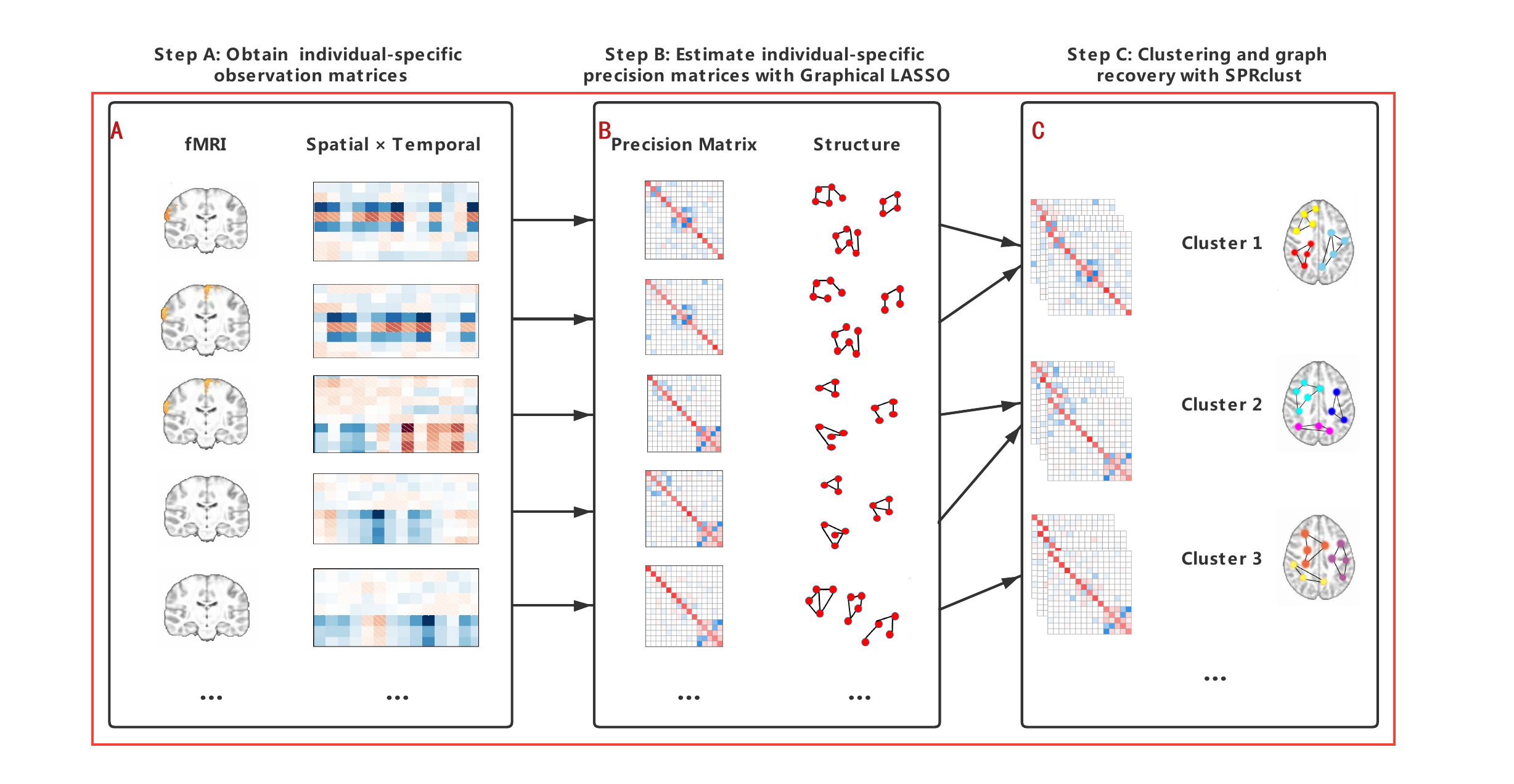}}
	\caption{Workflow of the SCEHG method: (A) Obtain the preprocessed Spatial $\times$ Temporal fMRI data for each individual; (B) Estimate the individual-specific precision matrices with Graphical LASSO ; (C) Clustering and Graph Recovery by SPRclust simultaneously.
	}
	\label{fig:workflow}
\end{figure}

Functional magnetic
resonance imaging (fMRI)  has been the mainstream imaging modality in neuroscience research. It records the blood oxygen level dependent time series.  fMRI data are in matrix-form with spatial (brain regions) by temporal (time points)  structure,  see Figure \ref{fig:workflow} (A). There also has been many literature on graphical modeling for matrix-valued data such as fMRI data, see for example, \cite{leng2012sparse,Zhou2014Gemini,Xia2017Hypothesis}.
In neuroscience studies, it is often the case that observations are from several distinct subpopulations/groups. A naive approach is to learn a graphical model for each group separately/independently. However, such a method inevitably ignores the common structure shared across different groups and is thus less inefficient and  suboptimal. To fully excavate the common structure across groups, series of joint estimation methods have been proposed during the last decade, see for example, \cite{guo2011joint,danaher2014joint,cai38joint,He2017Joint} for vector-valued data and \cite{Zhu2018multiple} for matrix-valued data.

All aforementioned literature crucially assume that the group label of each
observation is given/known in priori, which maybe not the case in real application such as online advertising (an important task of which is to target advertising better for a given user in an  online context with group label of each user unknown).  Clustering analysis is an essential tool for unsupervised machine learning to identify groups of objects of the similar pattern. With dissimilarity
structure  defined for each pair of objects (users in  the context of online advertising), many well-known clustering algorithms such as K-means,
K-medoids, hierarchical clustering can be applied.  One serious limitation of these traditional methods is the non-convexity of the corresponding optimization problems.  In the high-dimensional setting, clustering analysis with feature selection have drawn increasing attentions, see for example, \cite{Pan2007Penalized,Li2021Simultaneous}.

In many real applications, in addition to  grouping objects of the similar pattern, one may also be interested in understanding conditional dependence relationships among object attributes, which in turn helps to improve grouping/clustering accuracy.
The  methods for joint estimation of multiple graphs mentioned above are not directly applicable
as all those methods require that the group membership of each object  is known in advance. One may of course first do clustering analysis and then estimate the graphs of each ``clustered" group with memberships of objects obtained from the clustering analysis. However, the estimation accuracy of graph structures by this naive method heavily relies on the performance of different clustering algorithms and can not deal with the case that the number of underlying  clusters grows with the sample size.
Thus it's urgent to propose a method which can simultaneously conduct object clustering and joint graphical model estimation, especially in the era of big data. In this paper, we propose a method which can achieve Simultaneous Clustering and Estimation of Heterogeneous Graphs (SCEHG) for matrix-valued fMRI data. The typical characteristic of fMRI data is that it's in matrix-form and objects from different groups share the same zero  mean matrix but different covariance structures due to the the centralization preprocess step \citep{Chen2021Simultaneous}. Thus the clustering methods which rely on the mean difference of  groups do not work for fMRI data. Also, the estimation methods of graphical models for vector-valued data do not work (well) for fMRI data.
To overcome the challenge, we first construct individual-level between-region network measures for each subject by assuming a Kronecker product covariance matrices framework for fMRI data, see Figure \ref{fig:workflow} (B).  We then formulate the unsupervised clustering task as a supervised penalized regression learning task with grouping/fusion and sparsity penalty. The  fusion penalty  is for the purpose of clustering and sparsity penalty is for recovering  sparse graph structures  for heterogeneous groups.  A MDC-ADMM algorithm is proposed to solve the corresponding optimization problem and a generalized criterion to specify the number of clusters is also proposed. Both simulation study and real fMRI data analysis example show the superiority of the proposed SCEHG method. As a by-product of
this new method, we make R package SCEHG implementing MDC-ADMM algorithm available at GitHub  at \url{https://github.com/heyongstat/SCEHG}.

\subsection{Closely Related Literature  Review}
In the literature, a closely related line of research focuses on  clustering with fusion penalties  which apply to all the pairwise differences of centroids, and  these methods are typically known as regression/fusion-based clustering methods, see for example, \cite{Pan2013Cluster,Wu2016A,Zhang2019ROBUST}. Regression/fusion-based clustering methods show an advantage in some complex clustering situations such as when non-convex clusters exist, in which traditional clustering methods  $K$-means break down  \citep{Pan2013Cluster}.
The current work follows this line of research, but considers an additional sparsity penalty in the optimization problem to achieve variable selection.
Another closely related line of research focuses on both graphical model and clustering at the same time. \cite{Zhou2009Penalized} proposed a regularized Gaussian mixture model with general covariance matrices, which shrinks the means and covariance matrices and achieves  clustering and variable selection simultaneously. \cite{Gao2016Estimation} considered joint estimation of multiple precision matrices under the framework of Gaussian mixture models, but did
not enforce sparsity of  cluster means.  \cite{Hao2018Simultaneous} proposed a general framework of Simultaneous Clustering And estimatioN of heterogeneous graphical models (SCAN),  which is a likelihood-based method and treats the underlying group label as a latent variable. The literature on this line mentioned above all deals with vector-valued data, but not for matrix-valued data such as fMRI data, thus ignores the matrix-structure of fMRI data. Statisticians have recognized that ignoring the matrix-structure  gives rise to efficiency loss  in graphical modelling \citep{Zhu2018multiple,Ji2020Brain},   factor analysis \citep{wang2019factor,Yu2021Projected,He2021Vector} and discriminant analysis \citep{Hu2020Matrix}.
The last closely related line of research focuses on penalized model-based clustering of fMRI data. Fewer studies  focused on
clustering of subjects based on functional brain connectivity patterns. \cite{Zeng2014Unsupervised} proposed an unsupervised maximum margin clustering method and distinguished depressed patients from healthy controls based on functional connectivity estimates obtained from pairwise correlations, which cannot yield interpretable estimates of functional connectivity network structures for each class.
\cite{Dilernia2021Penalized}  proposed a random covariance clustering
method to simultaneously cluster subjects and obtain sparse precision matrix estimates for
each cluster as well as each subject, by which they also distinguished  schizophrenia patients from healthy controls.  However,  the mixture Wishart distribution assumption of the subject-level precision matrix may deviate from the reality and the assumption of independent observations within each subject is  violated for fMRI data which exhibit autocorrelation .

\subsection{Contributions and Structure of the Paper}
First we propose a general framework for clustering and variable selection in Section \ref{sec:2}. We used the non-convex penalty in \cite{Pan2013Cluster} for grouping pursuit which enforces the equality among some unknown subsets of parameter estimates. Different from \cite{Pan2013Cluster}, we add a LASSO penalty to encourage the sparsity of parameter estimates. We will see that the LASSO penalty is indispensable in terms of recovering sparse graphs of heterogeneous groups with fMRI data in Section \ref{sec:3}. We name the clustering method as \textit{sparse penalized regression-based clustering} (\textbf{SPRclust}).
For the corresponding optimization problem, we also combine difference of convex (DC) programming with the alternating direction method of multipliers (ADMM) as in  \cite{Pan2013Cluster}, but modifies the ADMM algorithm to further deal with the additional $L_1$ penalty part and thus we named the algorithm as MDC-ADMM (with the first M abbreviated for Modified). We prove the convergence of the MDC-ADMM algorithm and also propose a new criterion to select the tuning parameters which takes both the clustering performance and variable selection performance into account.

In Section \ref{sec:3} we introduce how we achieve Simultaneous Clustering and Estimation of Heterogeneous Graphs (SCEHG) of fMRI data with the SPRclust method presented in Section \ref{sec:2}, see also Figure \ref{fig:workflow} (C). We will see that the proposed procedure allows for the serial correlation of observations within each subject, thus fits to the analysis of  fMRI data better compared with the independent assumption in \cite{Dilernia2021Penalized}. Our method also serves as the first (as far as we know) clustering method based on covariance structure difference rather than mean difference of various groups, which is particularly suitable for fMRI data as the preprocessed fMRI data are usually demeaned. The SCEHG takes the serial autocorrelation of fMRI data into account when constructing the individual region-by-region connectivity strength and thus relaxes the i.i.d. assumptions in \cite{Chen2021Simultaneous,Dilernia2021Penalized}.

 Extensive simulation studies in Section \ref{sec:4} show the superiority of the proposed SCEHG method over some state-of-the-art methods. In Section \ref{sec:5} we analyzed a real fMRI dataset associated with ADHD by the SCEHG method and the findings are consistent with existing literature. We discuss the limitations of SCEHG method and possible future directions in Section \ref{sec:6}.

To end this section, we introduce some notations used throughout the study. Let $a_{+}$ denotes the positive part of $a$. { For a vector $\bmu=(\mu_1,\ldots,\mu_p)^\top\in \RR^{p}$, let $||\bmu||_1=\sum_{i=1}^p|\mu_i|$ and $||\bmu||_2=\sqrt{\sum_{i=1}^p\mu_i^2}$. For a matrix $\bX_{n\times p}$, let $\bX_{\cdot j}$ be the $j$-th column of $\bX$ and $X_{ij}$ be its $(i,j)$-th element. Denote $\bI_p$ be the $p$-dimensional identity matrix.} Furthermore, we denote the trace of $\bX$ as $\text{Tr}(\bX)$ and  the determinant of $\bX$ as $|\bX|$. Let $||\bX||_1$ be the $\ell_1$ norm, that is the sum of the absolute values of all the elements of $\bX$. We denote by $\mathrm{Vec}(\bX)$  the vector obtained by stacking the columns of $\bX$. Let $\mathrm{Vec}(\bX)_{j>i}$ be the operator that stacks the columns of the upper triangular elements of matrix $\bX$ excluding the diagonal elements to a vector. The notation $\otimes$ represents Kronecker product. For a set $\cF$, denote by $\sharp{\{\cF\}}$ the cardinality of $\cF$.

\section{A General Framework for Clustering and Variable Selection}\label{sec:2}
In this section, we introduce a general framework for achieving clustering and  parameter estimate sparsity simultaneously.   In Section \ref{sec:2.1}, we introduce the model setup with the optimization problem. In Section \ref{sec:2.2}, we introduce the MDC-ADMM algorithm for solving the optimization problem and study its convergence property.   In Section \ref{sec:2.3}, we propose a new criterion for tuning parameter selection.
\subsection{Model Setup}\label{sec:2.1}
Given dataset $\bX_{n\times p}=(\bx^{\top}_1,\bx^{\top}_2,\cdots,\bx^{\top}_n)^{\top}$ with  observations $\bx_{i}=(x_{i1},x_{i2},\cdots,x_{ip})^{\top}, i=1,\ldots,n$, we aim to conduct  cluster analysis to identify group-memberships of the observations such that the within-group similarity and between-group dissimilarity are  strong.

We assume that each data point~$\bx_i$~has its own centroid~$\bmu_i=(\mu_{i1},\mu_{i1},\cdots,\mu_{ip})^{\top}$, which can be its mean or median (or other measure), depending on the application. Our goal is to estimate~$\bmu_i$~while acknowledging the possibility that many~$\bmu_i$'s would be equal if their corresponding~$\bx_i$'s are from the same cluster. We adopt the fusion penalty to encourage the equality of the centroids. To alleviate the bias of the usual convex~$\ell_2$-norm group penalty, we adopt the non-convex
grouped truncated LASSO penalty as in \cite{Pan2013Cluster}, motivated by the Truncated $\ell_1$ Penalty  (TLP) function in \cite{shen2012likelihood}. TLP acts as the surrogate of $\ell_0$ penalty function and enjoys the advantages stated as ``adaptive model selection through adaptive shrinkage", ``piecewise linearity" and ``low resolutions" by \cite{shen2012likelihood}.
We also add the $\ell_1$-norm penalty of centroids to achieve sparsity of the centroids' estimates, which plays pivotal role in graph recovery introduced in Section \ref{sec:3}.

Summarizing the discussion above, we consider the following optimization problem:
\begin{equation}\label{equation:opt}
\begin{aligned}
  & \min _{\bmu, \btheta} \quad S(\bmu, \btheta)=\frac{1}{2} \sum_{i=1}^{n}\left\|\bx_{i}-\bmu_{i}\right\|_{2}^{2}+\lambda_1\sum_{i=1}^{n}\|\bmu_i\|_1+\lambda_2\sum_{i<j} \operatorname{TLP}\left(\left\|\btheta_{i j}\right\|_{2} ; \tau\right)\\
& \text { subject to }  \btheta_{i j}=\bmu_{i}-\bmu_{j}, \quad 1 \leq i<j \leq n.
\end{aligned}
\end{equation}
where~$\operatorname{TLP}(a,b)=\min(|a|,b)$ and $\lambda_1,\lambda_2$ are two tuning parameters which controls the sparsity and grouping effect, respectively.
By solving the above optimization problem, we obtain the estimates of the centroids, denoted as $\{\hat\bmu_i,1\leq i\leq n\}$. Then the observations with equal estimated centroids are naturally clustered together and  the estimated centroids show sparsity due to the first $\ell_1$ penalty in  (\ref{equation:opt}). The clustering method is named as \textit{sparse penalized regression-based clustering} (\textbf{SPRclust}). The sparsity of the estimated centroids is of great importance as it leads to the sparsity of the edges in heterogeneous graphs for fMRI data discussed in Section \ref{sec:3}. In the following we first discuss the computational aspects of the optimization problem.

 \subsection{Computational Aspects}\label{sec:2.2}
\subsubsection{Algorithm Implementation}

The optimization problem in (\ref{equation:opt}) is  non-convex on $\btheta_{ij}$ and can be similarly solved by a modified DC-ADMM as in \cite{Wu2016A}. Let~$S(\bmu,\btheta)=S_1(\bmu,\btheta)-S_2(\btheta)$~where
$$S_1(\bmu,\btheta)=\frac{1}{2} \sum_{i=1}^{n}\left\|\bx_{i}-\bmu_{i}\right\|_{2}^{2}+\lambda_1\sum_{i=1}^{n}||\bmu_i||_1+\lambda_2\sum_{i<j}||\btheta_{ij}||_1,\quad S_2(\btheta)=\lambda_2\sum_{i<j}(||\btheta_{ij}||_2-\tau)_+.$$

 Note that~$S_1$~and~$S_2$~are both convex functions and now $S(\bmu,\btheta)$ is decomposed into the difference of these two convex functions. We then construct a sequence of lower approximations of~$S_2(\btheta)$, namely $\{S^{(m)}_2(\btheta)\}$,
$$S^{(m)}_2(\btheta)=S_2(\hat{\btheta}^{(m)})+\lambda_2\sum_{i<j}(||\btheta_{ij}||_2-||\hat{\btheta}_{ij}^{(m)}||_2)I(||\hat{\btheta}_{ij}^{(m)}||_2 \geq \tau),$$
where $\hat{\btheta}_{ij}^{(m)}$ is the estimate from the $m$-th iteration.
Thus the corresponding~$S^{(m+1)}(\btheta,\bmu)$~can be given as

\begin{align}
S^{(m+1)}(\btheta,\bmu)=&\frac{1}{2} \sum_{i=1}^{n}\left\|\bx_{i}-\bmu_{i}\right\|_{2}^{2}+\lambda_1\sum_{i=1}^{n}||\bmu_i||_1+\lambda_2\sum_{i<j}||\btheta_{ij}||_2I(||\hat{\btheta}_{ij}^{(m)}||_2<\tau)\nonumber\\
&+\lambda_2\tau\sum_{i<j}I(||\hat{\btheta}_{ij}^{(m)}||_2\geq \tau).\nonumber
\end{align}

\begin{figure}[!h]
	\centerline{\includegraphics[width=16cm,height=8cm]{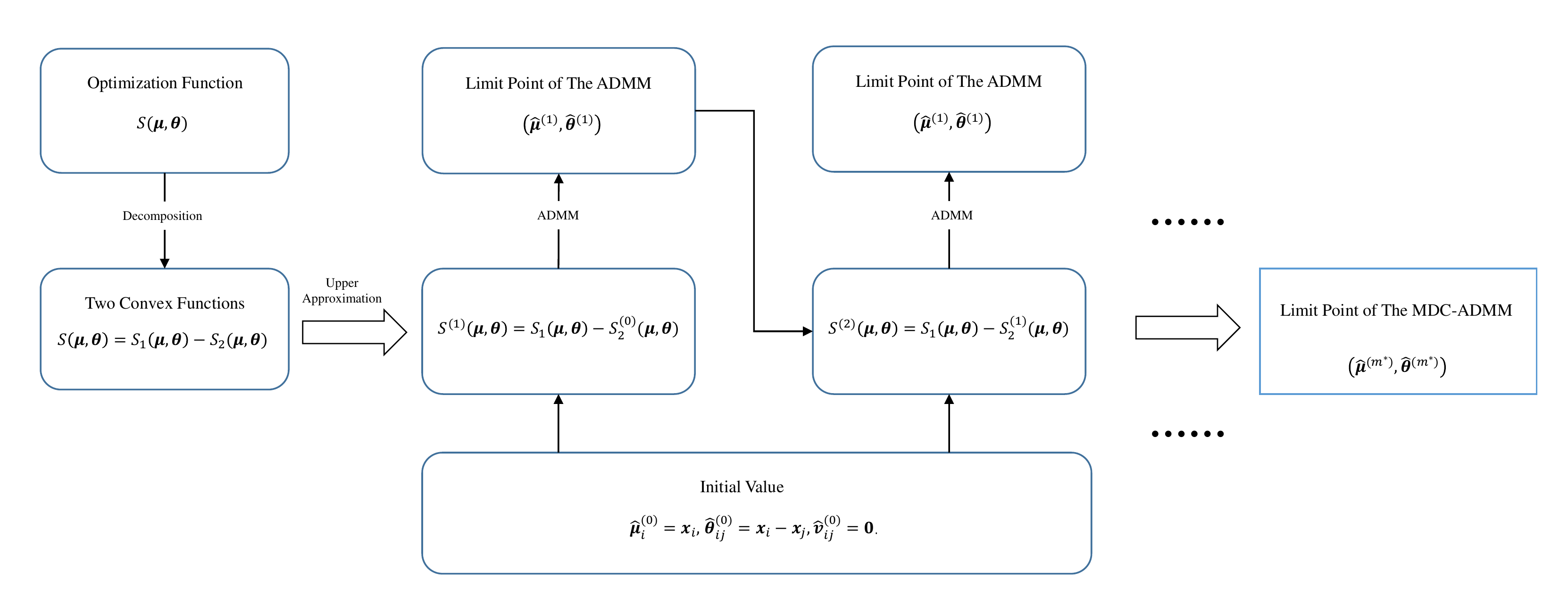}}
	\caption{Illustration of the MDC-ADMM Algorithm.
	}
	\label{fig:alworkflow}
\end{figure}

Apparently,~$S^{(m+1)}(\btheta,\bmu)$~is an upper approximation of $S(\btheta,\bmu)$ and it's convex over~$\btheta$~and~$\bmu$.
So~$(\ref{equation:opt})$~can be rewritten as
\begin{equation}\label{equation:2.2}
\min_{\btheta,\bmu}\quad S^{(m+1)}(\btheta,\bmu), \quad \text{subject to} \quad \btheta_{i j}=\bmu_{i}-\bmu_{j}, \quad 1 \leq i<j \leq n.
\end{equation}

Following~\cite{boyd2011distributed}, we form the scaled augmented Lagrangian function as
\begin{align}\label{equation:2.3}
L_{\rho}(\btheta,\bmu)= &\frac{1}{2} \sum_{i=1}^{n}\left\|\bx_{i}-\bmu_{i}\right\|_{2}^{2}+\lambda_1\sum_{i=1}^{n}||\bmu_i||_1+\lambda_2\sum_{i<j}||\btheta_{ij}||_2I(||\hat{\btheta}_{ij}^{(m)}||_2<\tau) \nonumber\\
&+\lambda_2\tau\sum_{i<j}I(||\hat{\btheta}_{ij}^{(m)}||_2\geq \tau)+\frac{\rho}{2}\sum_{i<j}||\btheta_{ij}-(\bmu_i-\bmu_j)+\bv_{ij}||^2_2-\frac{\rho}{2}\sum_{i<j}||\bv_{ij}||^2_{2},
\end{align}
where~$\bv_{ij}=\by_{ij}/\rho$, $\by_{ij}$~is the dual variable. The parameter $\rho$  affects the speed of convergence \citep{boyd2011distributed,Wu2016A} and we set it as 0.4 in our simulation study. Then we perform standard ADMM procedure as
\begin{align}\label{equation:2.4}
& \hat{\bmu}_{i}^{k+1}&=&\underset{\bmu_{i}}{\argmin}
\frac{1}{2}||\bx_i-\bmu_i||_2^2+\lambda_1||\bmu_i||_1+\frac{\rho}{2}\sum_{j>i}||\hat{\btheta}_{ij}^k-(\bmu_i-\hat{\bmu}_j^k)+\hat{\bv}^k_{ij}||^2_2\\
&&&+\frac{\rho}{2}\sum_{j<i}||\hat{\btheta}_{ji}^k-(\hat{\bmu}_j^{k+1}-\bmu_i)+\hat{\bv}^k_{ji}||^2_2,\nonumber\\
& \hat{\btheta}_{ij}^{k+1}&=&\underset{\btheta_{ij}}\argmin\left\{\begin{array}{ll}
\lambda_2\tau+\frac{\rho}{2}||\btheta_{ij}-(\hat{\bmu}_{i}^{k+1}-\hat{\bmu}_{j}^{k+1})+\hat{\bv}_{ij}^k||_2^2, \quad\quad\quad & \text{if}~||\hat{\btheta}_{ij}^{(m)}||_2\geq \tau;\\
\lambda_2||\btheta_{ij}||_2+\frac{\rho}{2}||\btheta_{ij}-(\hat{\bmu}_{i}^{k+1}-\hat{\bmu}_{j}^{k+1})+\hat{\bv}_{ij}^k||_2^2, & \text{if}~||\hat{\btheta}_{ij}^{(m)}||_2<\tau;\nonumber\\
\end{array}\right.\\
& \hat{\bv}_{ij}^{k+1}&=&\hat{\bv}_{ij}^{k}+\hat{\btheta}_{ij}^{k+1}-(\hat{\bmu}_{i}^{k+1}-\hat{\bmu}_{j}^{k+1}), \quad\quad 1\leq i<j\leq n,\nonumber
\end{align}
where superscript~$k+1$~means the $(k+1)$-th step~of ADMM iterations.

For the optimization in (\ref{equation:2.4}) to get the estimate  $\hat{\bmu}_{i}^{k+1}$,~we can construct pseudo observations~$(\bz_i^*,\by^*_i)$~to reformulate the problem as a standard LASSO problem. The first~$p$~rows of~$(\bz_i^*,\by^*_i)$~correspond to~$(\bI_{p},\bx_i)$~and the rest rows are constructed according to the last two terms of  equation $(\ref{equation:2.4})$. Specifically, we construct the pseudo observations $(\bz_i^*,\by^*_i)$ as follows:

$$(\bz_1^*,\by^*_1)=
\sqrt\frac{\rho}{2}\begin{pmatrix}
\frac{1}{\sqrt{\rho}}\bI_{p}  & \frac{1}{\sqrt{\rho}}\bx_1\\
\bm{\cZ}_2(1)  & \bm{\cY}_2(1)\\
\end{pmatrix},
(\bz_n^*,\by^*_n)=\sqrt\frac{\rho}{2}\begin{pmatrix}
\frac{1}{\sqrt{\rho}}\bI_{p}  & \frac{1}{\sqrt{\rho}}\bx_n\\
\bm{\cZ}_1(n)  & \bm{\cY}_1(n)\\
\end{pmatrix},
$$
and
$$(\bz_i^*,\by^*_i)=
\sqrt\frac{\rho}{2}\begin{pmatrix}
\frac{1}{\sqrt{\rho}}\bI_{p}  & \frac{1}{\sqrt{\rho}}\bx_i\\
\bm{\cZ}_1(i)  & \bm{\cY}_1(i)\\
\bm{\cZ}_2(i)  & \bm{\cY}_2(i)\\
\end{pmatrix}, 1<i<n,
$$
where $$(\bm{\cZ_1}(i),\bm{\cY_1}(i))=\begin{pmatrix}
-\bI_{p} & \hat{\btheta}_{1i}^k-\hat{\bmu}_1^{k+1}+\hat{\bv}^k_{1i}\\
-\bI_{p} & \hat{\btheta}_{2i}^k-\hat{\bmu}_2^{k+1}+\hat{\bv}^k_{2i}\\
\vdots              & \vdots\\
-\bI_{p} & \hat{\btheta}_{(i-1)i}^k-\hat{\bmu}_{(i-1)}^{k+1}+\hat{\bv}^k_{(i-1)i}\\
\end{pmatrix}, \quad
(\bm{\cZ_2}(i),\bm{\cY_2}(i))=\begin{pmatrix}
\bI_{p} & \hat{\btheta}_{i(i+1)}^k+\hat{\bmu}_{(i+1)}^k+\hat{\bv}^k_{i(i+1)}\\
\bI_{p} & \hat{\btheta}_{i(i+2)}^k+\hat{\bmu}_{(i+2)}^k+\hat{\bv}^k_{i(i+2)}\\
\vdots              & \vdots\\
\bI_{p} & \hat{\btheta}_{in}^k+\hat{\bmu}_{n}^k+\hat{\bv}^k_{in}\\
\end{pmatrix}.
$$
Thus, to update $\hat{\bmu}^{k+1}_i$~is equivalent to solving the following optimization problem:
\begin{equation}\label{equaiton:2.5}
\hat{\bmu}_{i}^{k+1}=\underset{\bmu_{i}}{\argmin} \left\{\|\by^{*}_{i}-\bz^{*}_i\bmu_i\|_2^2+\lambda_1\|\bmu_i\|_1\right\},
\end{equation}
from which we can see it's a standard LASSO problem and we use the cyclic coordinate descent algorithm to solve it \citep{friedman2010regularization}.

Similar to the group LASSO optimization problem in \cite{yuan2006model}, we update~$\hat{\btheta}_{ij}^{k}$~by soft thresholding operator, that is
$$\hat{\btheta}_{ij}^{k+1}=\left\{\begin{array}{ll}
\hat{\bmu}_{i}^{k+1}-\hat{\bmu}_{j}^{k+1}-\hat{\bv}_{ij}^k, \quad\quad& \text{if}~||\hat{\btheta}_{ij}^{(m)}||_2\geq \tau;\nonumber\\
\textbf{prox}_{\lambda_2/\rho}(\hat{\bmu}_{i}^{k+1}-\hat{\bmu}_{j}^{k+1}-\hat{\bv}_{ij}^k), & \text{if}~||\hat{\btheta}_{ij}^{(m)}||_2< \tau;\nonumber\\
\end{array}\right.\\$$
where~$\textbf{prox}_s(\bt)=(1-s/||\bt||_2)_+\bt$.

As an illustration, the whole MDC-ADMM algorithm is summarized in Algorithm \ref{alg:1}, see also the workflow in Figure \ref{fig:alworkflow}.
\begin{algorithm}[hbtp]
\caption{~MDC-ADMM Algorithm}
\label{alg:1}
\begin{algorithmic}[1]
\Require Dataset~$\bX=\{\bx_{1},\cdots,\bx_{n}\}$~; tuning parameters~$\lambda_{1},\lambda_2,\tau$~and~$\rho$.
\State \textbf{Initialize: }Set~$m=0,\hat{\bv}^{(0)}_{ij}=\bm{0},\hat{\bmu}_{i}^{(0)}=\bx_i$~and~$\hat{\btheta}_{ij}^{(0)}=\bx_i-\bx_j$~for~$1\leq i<j\leq n$.
\While {$m=0$~or~$S\left(\hat{\bmu}^{(m)},\hat{\btheta}^{(m)}\right)-S\left(\hat{\bmu}^{(m-1)},\hat{\btheta}^{(m-1)}\right)<0$~}
			\State $m\gets m+1$
			\State Update~$\hat{\bmu}^{(m)}$~and~$\hat{\btheta}^{(m)}$~based on~$(\ref{equation:2.4})$~until the convergence of ADMM.
\EndWhile
\Ensure Estimated centroids~$\hat{\bmu}_1,\hat{\bmu}_2,\cdots,\hat{\bmu}_n$~and a assigned cluster label for each observation.
\end{algorithmic}
\end{algorithm}

\subsubsection{Algorithm Convergence}

In Algorithm~\ref{alg:1}, for each iteration~$m$ of the ADMM algorithm, $\bmu_i^{(0)}=\bx_i$~and~$\hat{\btheta}_{ij}^{(0)}=\bx_i-\bx_j$~for~$1\le i<j\le n$~are used as the starting values; $(\hat{\bmu}^{(m+1)},\hat{\btheta}^{(m+1)})$~is the limit point of the ADMM iterations, or equivalently, is a minimizer of the Lagrangian function in (\ref{equation:2.3}). $(\hat{\bmu}^{(m+1)},\hat{\btheta}^{(m+1)})$~is then exploited to update the objective function~$S^{(m+1)}(\bmu,\btheta)$~as a new approximation to~$S(\bmu,\btheta)$. We iterate the above process until the stopping criteria are met. We have the following theorem which guarantee the convergence of the MDC-ADMM algorithm.

\begin{theorem} \label{Convergence}
In MDC-ADMM, $S(\bmu,\btheta)$~converges in a finite number of steps; that is, there is a~$m^*<\infty$~with
$$S\left(\bmu^{(m)},\btheta^{(m)}\right)=S\left(\bmu^{(m^*)},\btheta^{(m^*)}\right)\quad\quad~\text{for}~m \ge m^*.$$
Moreover, $\left(\bmu^{(m^*)},\btheta^{(m^*)}\right)$~is a KKT point.
\end{theorem}

\begin{remark}
Although~ADMM~algorithm ensures a global minimizer as~$S^{(m)}(\bmu,\btheta)$~is closed, proper and convex, MDC-ADMM only guarantee a KKT point as a result of the nonconvexity of~$S(\bmu,\btheta)$. A variant DC algorithm by \cite{breiman1993deterministic} can give a global minimizer, but the drawback lies in its slow convergence speed.  We prefer the present version for its faster convergence for large-scale problems. Furthermore, MDC-ADMM may yield different KKT points with different starting values even for the same dataset and parameters. However, as shown in our simulation study, the MDC-ADMM algorithm with the proposed initial values performs well for our purpose.
\end{remark}

\subsection{Criterion for Selecting Tuning Parameters}\label{sec:2.3}
For the optimization problem in (\ref{equation:opt}), we have three tuning parameters to be determined, i.e.,  $\tau,\lambda_1$, and $\lambda_2$. In TLP penalty,~$\tau$~controls the tolerance of~$||\btheta_{ij}||_2$. In our paper, we  choose the  tuning parameters $\tau,\lambda_1,\lambda_2$ by grid search. For criteria of grid search,~\cite{Pan2013Cluster}~and~\cite{ghadimi2014optimal}~suggested  GCV or the stability-based criterion based on (adjusted) rand index. However, these criteria are not directly applicable in our case as they ignores measuring the variable selection performance of the additional LASSO penalty. This poses a great challenge for tuning parameters selection in our setting. To overcome the challenge, we, as far as we know for the first time, propose a criterion to balance the performances of both clustering and variable selection for heterogeneous groups simultaneously. Our idea is motivated by the S4 criterion by \cite{Li2021Simultaneous}, abbreviated for ``Subsampling Score incorporating
Sensitivity and Specificity". The main idea of the S4 criterion is that the more stably the method performs, the better the tuning parameters are. It has achieved success in yielding good performance by measuring the stability of clustering in repeated subsampled data. In the following, we introduce our criterion for selecting $(\lambda_1,\lambda_2,\tau)$ in detail.

For each candidate tuning parameter combination ~$(\lambda_1,\lambda_2,\tau)$, we first obtain the estimated centroids of observations $\bX_{n\times p}$ by solving (\ref{equation:opt}) and the corresponding clustering result. We  then use a matrix~$\bT_{n\times n}=(T_{ij})$~to record the clustering result where~$T_{ij}$~indicates whether sample~$i$~and sample~$j$~are  in the same cluster:~$T_{ij}=1$~if samples~$i$~and~$j$~belong to the same cluster and~$0$~otherwise. Then we generate~$B$~sets of subsampled data, denoted as~$\bX_{[r\cdot n]\times p}^{(1)},\bX_{[r\cdot n]\times p}^{(2)},\cdots,\bX_{[r\cdot n]\times p}^{(B)}$, where $r\in(0,1)$ is the resampling fraction. Similarly, we can obtain~$\bT_{n\times n}^{(b)}=(T^{(b)}_{ij})$ for $b=1,2,\cdots,B$. Missing value label ~``NA"~is assigned to~$T^{(b)}_{ij}$~if one or both of the two samples~$i$~and~$j$~are not in the $b$-th subsampling dataset. We take element-wise average of~$\bT^{(b)}$~to derive the mean comembership matrix~$\bar{\bT}^{sub}=\bar{T}^{sub}_{ij}$, where~$\bar{T}^{sub}_{ij}$~indicates the frequency that sample~$i$~and sample~$j$~are clustered together across all~$B$~subsampling procedure. Missing values are omitted when taking average. The concordance score of sample $i$ is defined as
$$C_i=\frac{\sum_{j\ne i}\bar{T}^{sub}_{ij}I(T_{ij}=1)}{\sum_{j\ne i}I(T_{ij}=1)}+\frac{\sum_{j\ne i}(1-\bar{T}^{sub}_{ij})I(T_{ij}=0)}{\sum_{j\ne i}I(T_{ij}=0)}-1,~i=1,2,\cdots,n$$
where~$I(\cdot)$~is the indicator function. Assuming that the comembership matrix~$\bT$~is the underlying truth, the first term can be viewed as sensitivity
score of sample $i$ and the second term  as specificity score of sample~$i$. In our definition of~$C_i$, it is close to~$1$~when sample~$i$~is a stably clustered subject but approaches~$0$~if sample~$i$~is an outlier. We then truncate the lower~$\alpha$\%~of~$C_i$~to avoid the impact of potential outliers and define~$\bar{C}$~as the trimmed mean of~$C_i$.

The concordance score of features can be defined similarly as follows. As the nonzero locations of the centroids  may differ a lot across clusters, we consider the concordance score of features cluster by cluster. We first resort to the estimates of the centroids, $\hat\bmu_i=(\hat\mu_{i1},\ldots,\hat\mu_{ip})^\top$ to calculate~$\bm{f}_{k}=(f_{kj})$, where $j=1,\ldots,p$, $k=1,2,\cdots,\hat{K}$, where $\hat K$ is the corresponding estimated number of clusters. We set $f_{kj}=I\left((\sum_{i\in\cC_k}I(\hat{\mu}_{ij}\ne0)/|\cC_k|)>0.5\right)$, where $\cC_k$ is the label set of samples collected in the $k$-th cluster.
Similarly, $\bm{f}^{(b)}_{k}$~can be obtained by~$B$~times resampling and~$\bar{f}^{sub}_{kj}=(\sum_{b=1}^{B}f_{kj}^{(b)})/B$~is the proportion that feature~$j$~is selected among~$B$~times subsampling procedure. Then the concordance score of features in the ~$k$-th cluster can be defined as
$$F(k)=\frac{\sum_{j=1}^{p}\bar{f}^{sub}_{kj}I(f_{kj}=1)}{\sum_{j=1}^{p}I(f_{kj}=1)}+\frac{\sum_{j=1}^{p}(1-\bar{f}^{sub}_{kj})I(f_{kj}=0)}{\sum_{j=1}^{p}I(f_{kj}=0)}-1,$$
If the estimated number of clusters for~$\bX^{(b)}_{[r\cdot n]\times p}$~is larger than~that for $\bX_{n\times p}$, we omit this resampling dataset since~$F$~is not well defined.
For simplicity, we take the average of~$F(k)$ over $k$, denote as~$\bar{F}$~, to represent the concordance score of features corresponding to the given~$(\lambda_1,\lambda_2,\tau)$. Then our criterion to select tuning parameters can be summarized as follows:

1. Calculate~$\bar{C}$~and~$\bar{F}$~for each possible combination of $(\lambda_1,\lambda_2,\tau)$.

2. Choose the combinations whose~$\bar{C}$~are among the top $s\%$ and denote the set as~$\cA$~.

3. Choose the optimal combination~$(\lambda_1,\lambda_2,\tau)_{opt}$~which corresponds to the maximum~$\bar{F}$~in~$\cA$~, i.e., $(\lambda_1,\lambda_2,\tau)_{opt}=\argmax_{(\lambda_1,\lambda_2,\tau)\in \cA}\bar{F}$.\\

\begin{remark}
We always omit the cases when  all features are selected together or $\hat K=1$, as in these cases the denominator of the second term of $C_i$ or $F(k)$ (specificity) is zero and the corresponding
score is not well-defined. Our criterion first guarantees the performance of clustering and then guarantees the performance of variable selection. There are other ways to select the best tuning parameters based on~$\bar{C}$~and~$\bar{F}$~such as~$(\lambda_1,\lambda_2,\tau)_{opt}=\bargmax_{\lambda_1,\lambda_2,\tau}(\omega_1\bar{C}+\omega_2\bar{F})$~or~$(\lambda_1,\lambda_2,\tau)_{opt}=\bargmax_{\lambda_1,\lambda_2,\tau}\sqrt{(\omega_1\bar{C}\times\omega_2\bar{F})}$~where~$0<\omega_i<1$~and~$\omega_1+\omega_2=1$. The performances of these criteria are comparable by our simulation study. Throughout this paper, we set~$r=0.5$, $s\%=0.4$, $\alpha\%=0.2$ and~$B=5$~in both simulation study and real data analysis.

\end{remark}

\section{Simultaneous Clustering and Estimation of Heterogeneous Graphs with fMRI Data}\label{sec:3}
In this section, we introduce the SCEHG method that simultaneously conducts clustering and estimation of heterogeneous graphical models for matrix-variate fMRI data. We first review the definition of matrix-normal distribution  for characterizing the distribution of matrix-variate fMRI data. The framework is scientifically plausible in neuroimaging studies, see for example \cite{Xia2017Hypothesis,Zhu2018multiple,Chen2021Simultaneous}
\begin{definition}
A matrix-variate $\bZ_{p\times q}$ follows the matrix-normal distribution, denoted as
$$\bZ_{p\times q}\sim \cM\cN(\bM_{p\times q},\bSigma_T\otimes \bSigma_S),$$
if and only if $\mathrm{Vec}(\bZ_{p\times q})$ follows a multivariate normal distribution i.e.
$$\mathrm{Vec}(\bZ_{p\times q})\sim \cN(\mathrm{Vec}(\bM_{p\times q}),\bSigma_T \otimes \bSigma_S),$$
where $\bSigma_S=(\Sigma_{S,ij})\in \RR^{p\times p}$ and $\bSigma_T=(\Sigma_{T,ij})\in \RR^{q\times q}$ denotes the covariance matrices of $p$ spatial locations
and $q$ times points, respectively.
\end{definition}
By the matrix-normal framework,  we have $\text{Cov}^{-1}(\mathrm{Vec}(\bZ_{p\times q}))=\bSigma_T^{-1}\otimes \bSigma_S^{-1}=\bOmega_T\otimes \bOmega_S$, where $\bOmega_S$ denote the spatial precision matrix and $\bOmega_T$ the temporal precision matrix. The primary interest is to recover the connectivity network characterized by the spatial precision matrix $\bOmega_S$ while the
temporal precision matrix $\bOmega_T$ is treat as a nuisance parameter.
\subsection{Individual-specific between-region connectivity measures}\label{sec:3.1}
In the section, we introduce the procedure to construct the individual-specific between-region connectivity measures by estimating the spatial precision matrix for each subject. The technique is growing popular recently, also known as constructing  ``functional connectivity network predictors" for each subject, see for example, \cite{Chen2021Simultaneous,WEAVER2021Single}. Both  work omit the serial dependence of fMRI data and simply take the individual sample covariance matrix  as the input of CLIME \citep{cai2011constrained} or Graphical LASSO \citep{yuan2007model} algorithm. In this article, in view of the serial dependence of fMRI data, we propose a nonparametric method to estimate the spatial covariance matrix of each individual.

 Assume that there exist $K$ clusters and let $\cC_k$ be the label set of samples collected in the $k$-th cluster and $n=\sum_{k=1}^K|\cC_k|$. The preprocessed fMRI data are demeaned and thus the first moment information is not helpful for clustering, and clustering methods based on mean difference are invalid such as K-means.  We assume that
$$\bZ^{\gamma_k}\sim\cM\cN (\zero,\bSigma_{T_k}\otimes \bSigma_{S_k}), \quad \gamma_k\in \cC_k, \quad k\in \{1,2,\ldots,K\},$$
and without loss of generality, assume the  diagonal elements of $\bSigma_{T_k}$ are ones. At each time point $t\in\{1,2,\cdots,q\}$, we have $\bZ^{\gamma_k}_{\cdot t}\sim\cN(\zero,\bSigma_{S_k})$.
If the class label set $\cC_k$ is known in advance, there exists many algorithms to estimate $\bSigma_S^{-1}$ in the high-dimensional setting such as CLIME  or Graphical LASSO.
In the current work we tackle  an unsupervised learning problem, i.e., $\cC_k$ is unknown. We aim to cluster the samples and recover the heterogeneous networks  between brain regions of $K$ groups simultaneously. We take the serial dependence of fMRI data into account to estimate the spatial covariance matrix for each sample $\bZ^{\gamma}\in \mathbb{R}^{p\times q}$ (note that there is no subscript $k$ in $\gamma$ henceforth). In detail,  we  use the kernel method to estimate the individual-specific spatial covariance matrix at time point $t$ as:
\begin{equation} \label{tve}
\hat{\bSigma}^{\gamma}_S(t)=\frac{\sum_{s}\omega_{st}\bZ^\gamma_{\cdot s}(\bZ^\gamma_{\cdot s})^{\top}}{\sum_{s}\omega_{st}},
\end{equation}
from which we can see that $\hat{\bSigma}^{\gamma}_S(t)$ is a weighted covariance matrix of $\bZ^\gamma_{\cdot s}$, with weights $\omega_{sj}=K(|s-j|/h_n)$ given by a symmetric nonnegative kernel over time.  We set $h_n=n^{1/3}$ and use Gaussian kernel in (\ref{tve}). Finally, we set $\hat{\bSigma}^{\gamma}_S=\sum_{t=1}^{q}\hat{\bSigma}^{\gamma}_S(t)/q$ and use the graphical LASSO to estimate the individual-specific precision matrix ${\bOmega}^\gamma_S$,
\begin{equation} \label{Glasso}
\hat{\bOmega}^\gamma_S=\argmin_{\bOmega}\{\mathrm{Tr}\left(\hat{\bSigma}^\gamma_S\bOmega\right)-\log|\bOmega|+\lambda||\bOmega||_1\}, \gamma\in\{1,2\ldots,n\}.
\end{equation}
where the tuning parameter $\lambda$ is chosen by cross validation.

\subsection{SCEHG Method}\label{sec:3.2}
In this section, we formally introduce our SCEHG method. First, we  straighten the upper triangular matrix of $\hat{\bOmega}_S^{\gamma}$  without diagonal elements and treat them as new features. Then we adopt the SPRclust method proposed in section \ref{sec:2} to achieve clustering and estimation of heterogeneous graphs simultaneously.
 In detail, let $\bx_{\gamma}=\mathbf{Vec}(\hat{\bOmega}_S^{\gamma})_{j>i}, \gamma=1,2,\cdots,n$ and  $\bx_{\gamma}$ is a $p(p-1)/2$ dimensional vector. Then we denote $\bA_{\gamma}\in\mathbb{R}^{p\times p}, \gamma=1,2,\cdots,n$ and solve the following optimization problem:
 \begin{equation}\label{equation:scehg}
\begin{aligned}
  & \min _{\bA_{\gamma},\theta} \quad S(\bmu, \btheta)=\frac{1}{2} \sum_{\gamma=1}^{n}\left\|\bx_{\gamma}-\mathbf{Vec}(\bA_{\gamma})_{j>i}\right\|_{2}^{2}+\lambda_1\sum_{\gamma=1}^{n}||\mathbf{Vec}(\bA_{\gamma})_{j>i}||_1+\lambda_2\sum_{s<t} \operatorname{TLP}\left(\left\|\btheta_{s t}\right\|_{2} ; \tau\right)\\
& \text { subject to }  \btheta_{st}=\mathbf{Vec}(\bA_{s})_{j>i}-\mathbf{Vec}({\bA}_{t})_{j>i}, \quad 1 \leq s<t \leq n.
\end{aligned}
\end{equation}
 Optimization problem (\ref{equation:scehg}) is in essence the same with  optimization problem (\ref{equation:opt}) and can be solved by the MDC-ADMM algorithm.
 The SCEHG method  takes advantage of the the SPRclust in the following two aspects: (I) the features of SPRclust in the fMRI setting are the individual-specific between-region connectivity measures, i.e., the functional connectivity network predictors \citep{Chen2021Simultaneous,WEAVER2021Single}, and  we cluster the samples by their second moment covariance information rather than the mean; (II) as SPRclust achieve sparsity of the estimated centroids, in the current setting, it means that the solutions of optimization problem (\ref{equation:scehg}), i.e., $\hat\bA_{\gamma}, \gamma=1,\ldots,n$ are sparse, from which we can recover the edges of graphs by the nonzero locations of $\hat\bA_{\gamma}$.

\section{Simulation Studies}\label{sec:4}
In this section, we conduct simulation studies to assess the performance of the proposed SCEHG  method  in terms of both clustering and graph recovery. We consider the following method for comparison:
PRclust method by \cite{Wu2016A}, SCAN method by \cite{Hao2018Simultaneous} and sparse $K$-Means (SKM) by \cite{witten2010framework}.
For the SKM method, we select the tuning parameters involved by the S4 criterion in \cite{Li2021Simultaneous}.
For the   SCAN method, an initial value for the number of clusters should be given in advanced and we simply set it as the true number of clusters.

For simplicity, we consider generating data from a matrix-normal distribution with different $\bSigma_{S_k}$~but the same~$\bSigma_T$~in all clusters, that is
$$\bZ^{\gamma_k}\sim\cM\cN (\zero,\bSigma_{T}\otimes \bSigma_{S_k}), \quad \gamma_k\in \cC_k, \quad k\in \{1,2,\ldots,K\}.$$
We set the true number of clusters as $K=3$~and~$|\cC_k|\in\{10,15\}$,~$p\in\{10,15\}$~and $q=100$.
The covariance matrices structures are introduced in detail below.

\begin{figure}[!h]
	\centerline{\includegraphics[width=18cm,height=12cm]{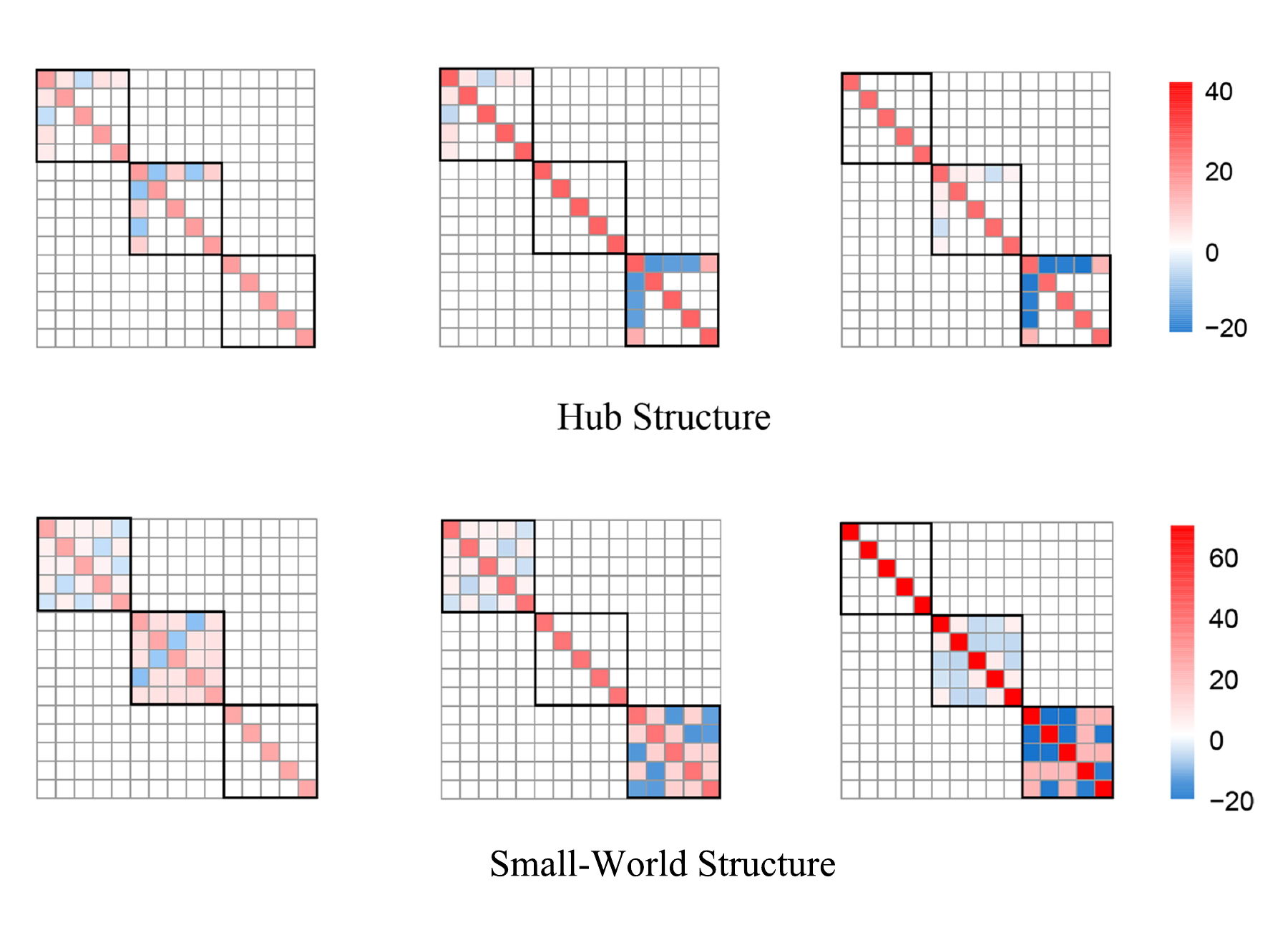}}
	\caption{Heat map of the generated precision matrices of $K=3$ clusters. The top panel illustrates the hub structure and the bottom panel illustrates the small-world structure.
	}
	\label{fig:heatmap}
\end{figure}

We consider  two types of $\bSigma_T$ which are common in fMRI data: the first  type is the auto-regressive (AR) correlation, where~$(\Sigma_{T})_{st}=0.5^{|s-t|}$; and the second  type is band correlation (BC), where~$(\Sigma_{T})_{st}=1/(|s-t|+1)$~for~$|s-t|<4$~and~$0$~otherwise.
As for $\bSigma_{S_k}$, we first introduce how we construct its inverse $\bOmega_{S_k}$, which characterizes the graph structures of $K$ groups.
We also consider two types of graph structure~$G_{S}$: the Hub structure and the Small-World structure. We resort to R package ``huge" to generate Hub structure with~$3$~non-overlapping graph and use R package ``rags2ridges" to generate~$3$~Small-World graph.  For further details of these two graph structures, one may refer to \cite{zhao2012huge} and \cite{Chen2021Simultaneous}. We design the precision matrices of $K=3$ groups such that all share the same graph structure (either Hub or Small-World) but with different partition blocks, illustrated by the heat map of the precision matrices in Figure \ref{fig:heatmap}.
The covariance matrices~$\bSigma_{S_k}$~are set to be~$\bOmega_{S_k}^{-1}$.

To conclude, we consider four scenarios according to the structures of $\bSigma_{S_k}$ and $\bSigma_T$:
\begin{itemize}
\item Scenario 1: $\bSigma_T$~is of AR covariance structure  and~$\bSigma_{S_k}$~is based on~$G_S$~with Hub structure.
\item Scenario 2: $\bSigma_T$~is of AR covariance structure  and~$\bSigma_{S_k}$~is based on~$G_S$~with Small-World structure.
\item Scenario 3: $\bSigma_T$~is of BD covariance structure  and~$\bSigma_{S_k}$~is based on~$G_S$~with Hub structure.
\item Scenario 4: $\bSigma_T$~is of BD covariance structure  and~$\bSigma_{S_k}$~is based on~$G_S$~with Small-World structure.
\end{itemize}

\begin{table}[htbp]
\centering
\caption{Simulation Results for Scenario 1.  Freq $a|b$,  $a$ and $b$ are the frequency of overestimating and underestimating the cluster numbers, respectively. The values in the parentheses denote standard deviation. }\label{Scenario1}
\scalebox{0.75}{
\begin{tabular}{ccccccccccc}
\toprule[2pt]
\multirow{2}{2cm}{\centering $(n_k,p,K)$}&\multirow{2}{2cm}{\centering Method}&\multicolumn{5}{c}{\centering Clustering-related Indexes}&\multicolumn{3}{c}{\centering Graph Recovery Indexes}\\
\cmidrule(lr){3-7} \cmidrule(lr){8-10}
&&$\hat{K}_{mean}$&Freq&Rand&aRand&Jaccard&TPR&TNR&FDR\\
\hline
\multirow{4}{2cm}{\centering $(10,10,3)$ }
&SCEHG           &2.8600 &0$|$14 &0.9678 &0.9375 &0.9404 &0.9026(0.1362) &0.8304(0.1321) &0.5600(0.1544)\\
&SKM                &2.9500 &4$|$10 &0.9686 &0.9356 &0.9347 &0.7750(0.3208) &0.4271(0.4824) &0.6486(0.3928)\\
&PRclust            &2.7600 &0$|$22 &0.9402 &0.8908 &0.9011 &0.9993(0.0115) &0.0284(0.0458) &0.8933(0.0134)\\
&SCAN &1.0000 &0$|$53 &0.3103 &0.0000 &0.3103 &1.0000(0.0000) &0.0061(0.0137) &0.8957(0.0013)\\
\hline
\multirow{4}{2cm}{\centering $(10,15,3)$ }
&SCEHG           &2.9800 &2$|$4  &0.9881 &0.9766 &0.9772 &0.8582(0.1506) &0.8207(0.0886) &0.6922(0.0835)\\
&SKM                &2.9800 &2$|$4  &0.9882 &0.9759 &0.9757 &0.6717(0.3829) &0.4546(0.4927) &0.6522(0.4346)\\
&PRclust            &2.9400 &0$|$6  &0.9862 &0.9732 &0.9745 &0.9929(0.0323) &0.0743(0.0801) &0.9182(0.0068)\\
&SCAN  &1.0000 &0$|$3  &0.3103 &0.0000 &0.3103 &1.0000(0.0000) &0.0447(0.0060) &0.9205(0.0005)\\
\hline
\multirow{4}{2cm}{\centering $(15,10,3)$ }
&SCEHG           &2.8600 &0$|$14 &0.9682 &0.9384 &0.9417 &0.9061(0.1371) &0.8339(0.1032) &0.5713(0.1344)\\
&SKM                &2.9100 &4$|$13 &0.9664 &0.9333 &0.9346 &0.6760(0.3454) &0.6022(0.4723) &0.5473(0.4324)\\
&PRclust            &2.8600 &0$|$14 &0.9682 &0.9384 &0.9417 &0.9993(0.0115) &0.0092(0.0195) &0.8954(0.0113)\\
&SCAN &1.0000 &0$|$52 &0.3182 &0.0000 &0.3182 &0.9971(0.0208) &0.0011(0.0049) &0.8965(0.0018)\\
\hline
\multirow{4}{2cm}{\centering $(15,15,3)$ }
&SCEHG           &2.9600 &1$|$5  &0.9885 &0.9777 &0.9787 &0.8742(0.1426) &0.8158(0.0943) &0.6925(0.0817)\\
&SKM                &2.9800 &2$|$4  &0.9894 &0.9789 &0.9788 &0.5725(0.3815) &0.6025(0.4820) &0.5636(0.4664)\\
&PRclust            &2.9300 &0$|$7  &0.9841 &0.9692 &0.9708 &0.9992(0.0102) &0.0324(0.0447) &0.9214(0.0036)\\
&SCAN &1.0000 &0$|$42 &0.3182 &0.0000 &0.3182 &0.9960(0.0180) &0.0129(0.0132) &0.9232(0.0017)\\
\bottomrule[2pt]
\end{tabular}
}
\end{table}

\begin{table}[htbp]
\centering
\caption{Simulation results for Scenario 2,  Freq $a|b$,  $a$ and $b$ are the frequency of overestimating and underestimating the cluster numbers, respectively. The values in the parentheses denote standard deviation.}\label{Scenario2}
\scalebox{0.75}{
\begin{tabular}{ccccccccccc}
\toprule[2pt]
\multirow{2}{2cm}{\centering $(n_k,p,K)$}&\multirow{2}{2cm}{\centering Method}&\multicolumn{5}{c}{\centering Clustering-related Indexes}&\multicolumn{3}{c}{\centering Graph Recovery Indexes}\\
\cmidrule(lr){3-7} \cmidrule(lr){8-10}
&&$\hat{K}_{mean}$&Freq&Rand&aRand&Jaccard&TPR&TNR&FDR\\
\hline
\multirow{4}{2cm}{\centering $(10,10,3)$ }
&SCEHG           &2.9800 &0$|$2  &0.9954 &0.9911 &0.9915 &0.8509(0.1463) &0.9156(0.0891) &0.2663(0.1787)\\
&SKM                &3.1200 &12$|$0 &0.9934 &0.9838 &0.9787 &0.7765(0.3389) &0.3536(0.4626) &0.6413(0.3464)\\
&PRclust            &2.9900 &2$|$3  &0.9927 &0.9856 &0.9859 &0.9825(0.0537) &0.0786(0.0965) &0.8106(0.0427)\\
&SCAN &1.0000 &0$|$54 &0.3103 &0.0000 &0.3103 &1.0000(0.0000) &0.0095(0.0168) &0.8208(0.0025)\\
\hline
\multirow{4}{2cm}{\centering $(10,15,3)$ }
&SCEHG           &3.0100 &1$|$0  &0.9998 &0.9995 &0.9993 &0.7138(0.1801) &0.8979(0.0799) &0.3335(0.1338)\\
&SKM                &3.0700 &7$|$0  &0.9973 &0.9935 &0.9914 &0.6973(0.4178) &0.3447(0.4695) &0.6428(0.3551)\\
&PRclust            &6.1900 &54$|$0 &0.9552 &0.8826 &0.8556 &0.8982(0.1311) &0.2497(0.2091) &0.7699(0.0549)\\
&SCAN  &1.0000 &0$|$5  &0.3103 &0.0000 &0.3103 &1.0000(0.0000) &0.0306(0.0158) &0.8047(0.0026)\\
\hline
\multirow{4}{2cm}{\centering $(15,10,3)$  }
&SCEHG           &2.9400 &0$|$6  &0.9864 &0.9736 &0.9750 &0.8298(0.1510) &0.9315(0.0747) &0.2305(0.1807)\\
&SKM                &3.1000 &10$|$0 &0.9948 &0.9876 &0.9838 &0.6504(0.3702) &0.5631(0.4714) &0.5338(0.3990)\\
&PRclust            &2.9600 &0$|$4  &0.9909 &0.9824 &0.9833 &0.9900(0.0398) &0.0444(0.0675) &0.8160(0.0370)\\
&SCAN &1.0000 &0$|$49 &0.3182 &0.0000 &0.3182 &0.9981(0.0132) &0.0107(0.0181) &0.8209(0.0033)\\
\hline
\multirow{4}{2cm}{\centering $(15,15,3)$ }
&SCEHG           &3.0100 &1$|$0  &0.9999 &0.9997 &0.9996 &0.7239(0.1871) &0.8923(0.0854) &0.3401(0.1368)\\
&SKM                &3.0400 &4$|$0  &0.9981 &0.9953 &0.9939 &0.6680(0.4250) &0.3833(0.4795) &0.6238(0.3663)\\
&PRclust            &6.1100 &61$|$0 &0.9666 &0.9175 &0.8951 &0.9326(0.1188) &0.1656(0.1904) &0.7845(0.0462)\\
&SCAN &1.0000 &0$|$18 &0.3182 &0.0000 &0.3182 &0.9981(0.0079) &0.0146(0.0113) &0.8075(0.0019)\\
\bottomrule[2pt]
\end{tabular}
}
\end{table}

\begin{table}[htbp]
\centering
\caption{Simulation results for Scenario 3, Freq $a|b$,  $a$ and $b$ are the frequency of overestimating and underestimating the cluster numbers, respectively. The values in the parentheses denote standard deviation.}\label{Scenario3}
\scalebox{0.75}{
\begin{tabular}{ccccccccccc}
\toprule[2pt]
\multirow{2}{2cm}{\centering $(n_k,p,K)$}&\multirow{2}{2cm}{\centering Method}&\multicolumn{5}{c}{\centering Clustering-related Indexes}&\multicolumn{3}{c}{\centering Graph Recovery Indexes}\\
\cmidrule(lr){3-7} \cmidrule(lr){8-10}
&&$\hat{K}_{mean}$&Freq&Rand&aRand&Jaccard&TPR&TNR&FDR\\
\hline
\multirow{4}{2cm}{\centering $(10,10,3)$ }
&SCEHG           &2.9100 &0$|$9  &0.9793 &0.9598 &0.9617 &0.9516(0.1003) &0.7056(0.1306) &0.7025(0.0942)\\
&SKM                &3.0200 &6$|$4  &0.9809 &0.9588 &0.9555 &0.8035(0.3132) &0.3596(0.4697) &0.6880(0.3770)\\
&PRclust            &2.7800 &0$|$21 &0.9471 &0.9008 &0.9080 &0.9993(0.0115) &0.0198(0.0324) &0.8943(0.0123)\\
&SCAN &1.0000 &0$|$58 &0.3103 &0.0000 &0.3103 &0.9948(0.0276) &0.0041(0.0109) &0.8964(0.0022)\\
\hline
\multirow{4}{2cm}{\centering $(10,15,3)$ }
&SCEHG           &2.9700 &1$|$4  &0.9883 &0.9771 &0.9779 &0.7309(0.1699) &0.9289(0.0536) &0.4956(0.1481)\\
&SKM                &3.0000 &4$|$4  &0.9877 &0.9747 &0.9734 &0.7058(0.3768) &0.4052(0.4863) &0.6817(0.4189)\\
&PRclust            &2.9500 &0$|$5  &0.9885 &0.9777 &0.9787 &0.9938(0.0291) &0.0564(0.0623) &0.9198(0.0049)\\
&SCAN  &1.0000 &0$|$2  &0.3103 &0.0000 &0.3103 &1.0000(0.0000) &0.0206(0.0146) &0.9223(0.0011)\\
\hline
\multirow{4}{2cm}{\centering $(15,10,3)$ }
&SCEHG           &2.9100 &0$|$9  &0.9795 &0.9604 &0.9625 &0.9155(0.1280) &0.8133(0.1043) &0.6001(0.1281)\\
&SKM                &2.9400 &3$|$9  &0.9781 &0.9568 &0.9579 &0.7030(0.3348) &0.5819(0.4759) &0.5585(0.4194)\\
&PRclust            &2.5500 &0$|$43 &0.8932 &0.7996 &0.8155 &1.0000(0.0000) &0.0045(0.0128) &0.8958(0.0109)\\
&SCAN&1.0000 &0$|$62 &0.3182 &0.0000 &0.3182 &0.9978(0.0169) &0.0021(0.0068) &0.8963(0.0018)\\
\hline
\multirow{4}{2cm}{\centering $(15,15,3)$ }
&SCEHG           &2.8900 &0$|$11 &0.9750 &0.9516 &0.9542 &0.8939(0.1308) &0.7612(0.1139) &0.7391(0.0772)\\
&SKM                &2.9800 &2$|$4  &0.9892 &0.9784 &0.9782 &0.5842(0.3810) &0.5924(0.4839) &0.5695(0.4612)\\
&PRclust            &2.9400 &0$|$6  &0.9864 &0.9736 &0.9750 &0.9996(0.0072) &0.0209(0.0301) &0.9223(0.0023)\\
&SCAN &1.0000 &0$|$12 &0.3182 &0.0000 &0.3182 &1.0000(0.0000) &0.0163(0.0082) &0.9226(0.0006)\\
\bottomrule[2pt]
\end{tabular}
}
\end{table}

\begin{table}[htbp]
\centering
\caption{Simulation results for Scenario 4,  Freq $a|b$,  $a$ and $b$ are the frequency of overestimating and underestimating the cluster numbers, respectively. The values in the parentheses denote standard deviation.}\label{Scenario4}
\scalebox{0.75}{
\begin{tabular}{ccccccccccc}
\toprule[2pt]
\multirow{2}{2cm}{\centering $(n_k,p,K)$}&\multirow{2}{2cm}{\centering Method}&\multicolumn{5}{c}{\centering Clustering-related Indexes}&\multicolumn{3}{c}{\centering Graph Recovery Indexes}\\
\cmidrule(lr){3-7} \cmidrule(lr){8-10}
&&$\hat{K}_{mean}$&Freq&Rand&aRand&Jaccard&TPR&TNR&FDR\\
\hline
\multirow{4}{2cm}{\centering $(10,10,3)$ }
&SCEHG          &2.9700 &0$|$3  &0.9931 &0.9866 &0.9872 &0.8519(0.1421) &0.9189(0.0640) &0.2760(0.1612)\\
&SKM                &3.1500 &14$|$0 &0.9923 &0.9810 &0.9752 &0.8054(0.3191) &0.3238(0.4523) &0.6560(0.3293)\\
&PRclust            &3.0300 &6$|$3  &0.9919 &0.9837 &0.9832 &0.9860(0.0454) &0.0634(0.0864) &0.8128(0.0411)\\
&SCAN &1.0000 &0$|$63 &0.3103 &0.0000 &0.3103 &0.9988(0.0093) &0.0109(0.0165) &0.8208(0.0030)\\
\hline
\multirow{4}{2cm}{\centering $(10,15,3)$ }
&SCEHG           &2.9800 &0$|$2  &0.9954 &0.9911 &0.9915 &0.7583(0.1741) &0.8261(0.1032) &0.4561(0.1190)\\
&SKM                &3.0900 &9$|$0  &0.9961 &0.9905 &0.9874 &0.7683(0.3827) &0.2749(0.4399) &0.6761(0.3211)\\
&PRclust            &3.0000 &0$|$0  &1.0000 &1.0000 &1.0000 &0.9257(0.1244) &0.1520(0.1399) &0.7945(0.0153)\\
&SCAN  &1.0000 &0$|$4  &0.3103 &0.0000 &0.3103 &1.0000(0.0000) &0.0235(0.0166) &0.8058(0.0027)\\
\hline
\multirow{4}{2cm}{\centering $(15,10,3)$ }
&SCEHG           &2.9300 &0$|$7  &0.9841 &0.9692 &0.9708 &0.8105(0.1512) &0.9491(0.0449) &0.2039(0.1505)\\
&SKM                &3.0800 &8$|$0  &0.9957 &0.9896 &0.9864 &0.7133(0.3574) &0.4852(0.4773) &0.5729(0.3775)\\
&PRclust            &2.8700 &0$|$13 &0.9705 &0.9428 &0.9458 &0.9904(0.0373) &0.0309(0.0549) &0.8185(0.0349)\\
&SCAN &1.0000 &0$|$64 &0.3182 &0.0000 &0.3182 &1.0000(0.0000) &0.0055(0.0146) &0.8214(0.0022)\\
\hline
\multirow{4}{2cm}{\centering $(15,15,3)$ }
&SCEHG           &2.9400 &0$|$6  &0.9864 &0.9736 &0.9750 &0.7397(0.1795) &0.8673(0.0965) &0.3877(0.1313)\\
&SKM                &3.0300 &3$|$0  &0.9988 &0.9972 &0.9963 &0.6940(0.4137) &0.3625(0.4734) &0.6333(0.3579)\\
&PRclust            &2.9600 &0$|$4  &0.9909 &0.9824 &0.9833 &0.9593(0.0940) &0.0895(0.1139) &0.8005(0.0133)\\
&SCAN &1.0000 &0$|$21 &0.3182 &0.0000 &0.3182 &1.0000(0.0000) &0.0190(0.0146) &0.8065(0.0023)\\
\bottomrule[2pt]
\end{tabular}
}
\end{table}

In the following we introduce related indexes to evaluate the performance of our SCEHG method in terms of both clustering and graph recovery. The Rand index, adjusted Rand index~(aRand) and Jaccard index are common indexes for evaluating the performance
of clustering methods, see for example, ~\cite{rand1971objective}, \cite{hubert1985comparing}~and~\cite{Wu2016A}.  The closer the values of these indexes are to 1, the better the clustering performances are.
We also report the mean of the estimated cluster numbers by various methods over 100 replications, denoted as ~$\hat{K}_{mean}$~and summarize the frequency of overestimating and underestimating the cluster numbers.
To evaluate the performance of graph recovery, we adopt the common indexes, the true positive rate~(TPR), true negative rate~(TNR)~and false discovery rate~(FDR) see for example, \cite{Chen2021Simultaneous}.

The detailed results are shown in Tables \ref{Scenario1}-\ref{Scenario4}, from which we can see that the proposed SCEHG performs satisfactorily in various scenarios. We can also see that the  SCEHG  show advantage over the other three methods. The SKM method and the PRclust method tend to perform not too badly (at most comparable with the  SCEHG  method) in terms of clustering but are much inferior to the proposed  SCEHG in terms of graph recovery. As PRclust cannot achieve sparsity of the estimated centroids, it always leads to high TPR but low TNR. The SCAN method seems to lose power in the designed setting and performs not satisfactorily in terms of both clustering and graph recovery. To conclude, our SCEHG method shows advantages over the existing state-of-the-art methods in terms of both clustering and graph structure recovery.

\section{Real Analysis of fMRI data related with ADHD }\label{sec:5}

 Attention Deficit Hyperactivity Disorder (ADHD), is one of the most common neurodevelopmental disorders in children. It is estimated that ADHD has a worldwide prevalence of 7.2\% among children, the condition persists into adulthood among 70\% of those diagnosed in childhood \citep{Zhao2017Abnormal}.
In this section, we apply the proposed SCEHG method to analyze a resting state fMRI dataset associated with ADHD. The dataset is from the ADHD-200
Global Competition, which includes demographical
information and resting-state fMRI of nearly one thousand
children and adolescents, including both combined types
of ADHD and typically developing controls (TDC). The
data were collected from eight participating sites and we focus
our analysis on the fMRI data from the Beijing site only to avoid potential site bias.  The dataset consists of 183 participants with 74 patients (ADHD) and 109 controls (TDC) and can be downloaded from \url{https://neurobureau.projects.nitrc.org}.  To reduce computational burden, we  only focus on  52 ADHD patients and 25 controls from the male participants and partition the standard 116 nodes into 8 brain regions: Insula, Limbic,  Occipital, Parietal, Frontal, Cerebellum, SCGM and Temporal. We then  take the average over each brain region at given time points for each subject, finally leading to spatial dimension $p=8$ , temporal dimension $ q=232 $  and sample size $ n=77 $ . For each subject, we use the algorithm proposed in Section \ref{sec:3.1} by the \textsf{R} package ``glasso"  to obtain the individual precision matrix. Then we use the  proposed SCEHG method to cluster the subjects and estimate the brain connectivity graphs.

\begin{table}[htbp]
	\centering
	\caption{The clustering results for ADHD and TDC samples.}\label{res}
	\scalebox{1}{
		\begin{tabular}{ccccccccccc}
			\hline
			&Group &ADHD &TDC &Total\\
			\hline
			&1 &49$(94.23\%)$ &2$(8\%)$  &51\\
			&2 &3$(5.77\%)$ &23$(92\%)$  &26\\
			\hline
			&Total &52 &25\\
			\hline
		\end{tabular}
	}

\end{table}

The clustering results of the samples are shown in Table \ref{res}, from which we can see that most of ADHD samples are clustered into Group 1, and over 90 percent of TDC samples are clustered into Group 2.  The Rand index and the adjusted Rand index  are 0.877 and 0.751, respectively. It has to be pointed out that the above results are based on a priori that the clustering number is two.  Actually, the subjects in Group 1 are indeed clustered into one group by the  SCEHG method , while the subjects in Group 2 are not exactly clustered into one group by  SCEHG method. Given the priori that $K=2$, the subjects that are not  clustered into Group 1 are assigned to Group 2 artificially. In fact, the TDCs may have other potential mental diseases and there is no guarantee that TDCs can be clustered into the same group by any clustering technique such as  $K$-means.

\begin{table}[htbp]
	\centering
	\caption{Proportion of the edges absent in Group 1 while existent in Group 2.}\label{eage}
	\scalebox{1}{
		\begin{tabular}{ccccccccccc}
			\toprule[2pt]
			&Edge & Group 1 & Group 2 \\
			\hline
			&Insula $\leftrightarrow$ Limbic  &100\% &46.2\% \\
			&Insula $\leftrightarrow$ Occipital &94.1\% &42.3\% \\
			&Insula $\leftrightarrow$ Parietal &94.1\% &30.8\%\\
			&Frontal $\leftrightarrow$ SCGM &92.2\% &42.3\%\\
			&Insula $\leftrightarrow$ Temporal &92.2\% &42.3\%\\
			&Limbic $\leftrightarrow$ Temporal &90.2\% &42.3\%\\
			&Occipital $\leftrightarrow$ Cerebellum &88.2\% &34.6\%\\
			&Frontal $\leftrightarrow$ Occipital &88.2\% &42.3\%\\
			&Limbic $\leftrightarrow$ Occipital &84.3\% &30.8\%\\
			&Limbic $\leftrightarrow$ Parietal &82.4\% &42.3\%\\
			&Frontal $\leftrightarrow$ Cerebellum &82.4\% &38.5\%\\
			&SCGM $\leftrightarrow$ Temporal &82.4\% &42.3\%\\
			&Frontal $\leftrightarrow$ Limbic &80.4\% &26.9\%\\
			&Insula $\leftrightarrow$ SCGM &74.5\% &38.5\%\\
			\bottomrule[2pt]
		\end{tabular}
	}

\end{table}

 Next we  analyze the resultant  brain connectivity graphs and identify the the connectivity
networks between brain regions. By the SCEHG Method proposed in Section \ref{sec:3.2}, we obtain the estimates of $\bA_{\gamma}$ for each subject, denoted as $\hat{\bA}_{\gamma},\gamma=1,2,\cdots,n$. Then the proportion  of an absence of edge between brain region $i$ and $j$ in Group $k$ ($k=1,2$) can be calculated by the following formula:
$$
\text{Prop}(e_{ij}^{(k)})=\dfrac{\sum_{\gamma \in \cC_k}I(\hat{\bA}_{\gamma,ij}=0)}{\sharp \left\{ \cC_k \right\} }
$$
where $\cC_k,k=1,2$ is the index set of Group $k$, $\hat{\bA}_{\gamma,ij}$ is the $(i,j)$-th entry of $\hat{\bA}_{\gamma}$. The connection between brain region $i$ and $j$ in Group $k$ is thought  to be absent if $\text{Prop}(e_{ij}^{(k)}) >=0.5$, otherwise there exist an edge between brain region $i$ and $j$. The proportions of the edges absent in Group 1 while existent in Group 2 are displayed in Table \ref{eage}, while all other edges are absent in two groups. From Table \ref{eage}, it can be deduced that the cause of ADHD may be attributed to the absent connections between certain brain regions.

\begin{figure}[!h]
	\centering
	\includegraphics[width=1\linewidth,height=6cm]{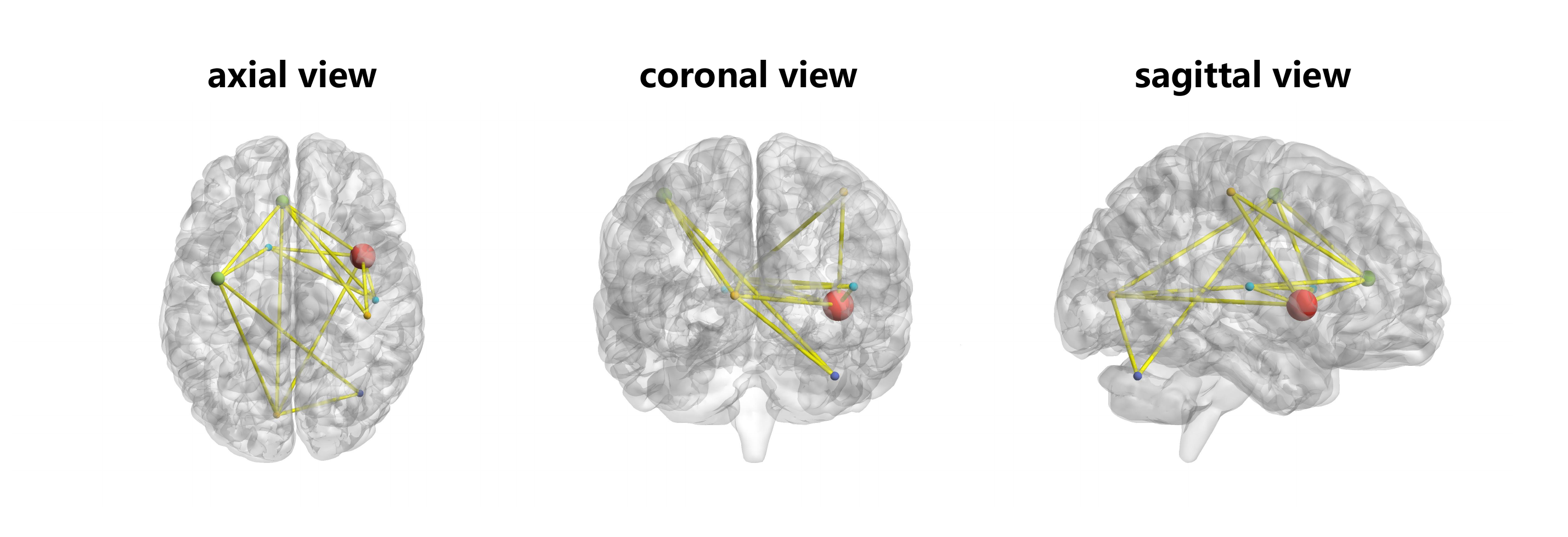}
	\caption{The differential network between Group 1 and Group 2 from the axial view, coronal view and sagittal view respectively (left ro right). The red point is ``Insula" brain region.}
	\label{fig:insula}
\end{figure}

Furthermore, we find that the brain region ``Insula" is a hub node in the differential network, which is shown in Figure \ref{fig:insula}. We suspect that the brain region ``Insula" is impaired in the ADHD patients as the functional connectivity between this region with other regions are almost all absent in ADHD patients. In fact it has been proved that the ``Insula" brain region is closely related to the pathophysiology of ADHD: Deficits of the anterior insula (AI) which
is involved in salient stimuli allocation might be associated with the pathophysiology of
ADHD according to \cite{Zhao2017Abnormal}, see also \cite{Vetter2018Anterior,Belden2014Anterior}.

\section{Summary and Discussions}\label{sec:6}
Graphical model or network model plays an important role in characterizing the relationship between variables and has been widely applied in areas such as biology and finance.  Most existing literature focuses on the case that the membership of each observation is known in priori, which may not hold in real application. For example, in neuroscience studies,  one usually does not known how many subtypes of a mental disease exist. In this article, we propose a method which can achieve clustering and graph recovery simultaneously for matrix-variate data, motivated by the fMRI technique.
In essence, we transform the unsupervised learning problem into a supervised penalized regression-based problem, with both $\ell_1$ penalty and fusion penalty. We propose a MDC-ADMM algorithm for the optimization problem. Both simulation study and real data analysis show the advantage of the proposed method in terms of both clustering and graph recovery. Limitations of the proposed method lie in its high computation burden when dealing with large-scale problem (large sample size and dimensionality) and its inability of controlling the resultant number of clusters,  which are common limitations for the fusion-penalty based regression technique for clustering. In future study we will consider  more efficient algorithms or methods to achieve clustering and graph recovery simultaneously for large-scale problem.

\section*{Acknowledgements}
This work was supported by grants from  the National Natural Science Foundation of China (Grant No.  12171282, 11801316, 11971116);  Natural Science Foundation of Shandong Province (Grant No. ZR2019QA002); the Fundamental Research Funds of Shandong University, China; Young Scholars Program of Shandong University, China.  We would like to thank professor
Will Wei Sun at Purdue University for providing the  codes of their SCAN method.

\appendix
\section{Appendix}
\textbf{Proof of Theorem 2.1~}
The  convergence of MDC-ADMM in finite steps attributes to the following three facts. First, as $L_{\rho}(\btheta,\bmu)$ in (\ref{equation:2.3})~is closed, proper and convex and the unaugmented Lagrangian $S^{(m+1)}(\btheta,\bmu)$ in (\ref{equation:2.2})~has a saddle point, thus the ADMM algorithm converges to the optimal value, see \cite{boyd2011distributed}. Second, by construction of ~$S^{(m)}(\bmu,\btheta)$~and~$S(\bmu,\btheta)$, for each positive integer $m$, we have
$$S\left(\bm{\hat{\mu}}^{(m)},\bm{\hat{\theta}}^{(m)}\right)= S^{(m+1)}\left(\bm{\hat{\mu}}^{(m)},\bm{\hat{\theta}}^{(m)}\right)\leq S^{(m)}\left(\bm{\hat{\mu}}^{(m)},\bm{\hat{\theta}}^{(m)}\right)\leq S^{(m)}\left(\bm{\hat{\mu}}^{(m-1)},\bm{\hat{\theta}}^{(m-1)}\right)=
S\left(\bm{\hat{\mu}}^{(m-1)},\bm{\hat{\theta}}^{(m-1)}\right),$$
which implies~$S(\bmu^{(m)},\btheta^{(m)})$~is non-increasing with respect to~$m$~and the equality is established if and only if ~$\left(\bm{\hat{\mu}}^{(m)},\bm{\hat{\theta}}^{(m)}\right)=\left(\bm{\hat{\mu}}^{(m-1)},\bm{\hat{\theta}}^{(m-1)}\right)$~as a result of the convergence of ADMM algorithm. Thus the monotonicity of~$S(\bmu^{(m)},\btheta^{(m)})$~with respect to~$m$~can be regarded as the stopping criterion. Third, as~$S^{(m+1)}(\bmu,\btheta)$~depends on $m$ only through the indicator function ~$I\left(||\btheta_{ij}^{(m)}||_2\ge \tau\right)$, which implies $S^{(m+1)}(\bmu,\btheta)$ has a finite set of possible functional forms across all $m$ and this leads to a finite set of distinct minimal values. These facts imply MDC-ADMM terminates in a finite number of iterations.

Next we show that~$\left(\bmu^{(m^*)},\btheta^{(m^*)}\right)$~is a KKT point of~$S(\bmu,\btheta)$. Following Theorem~$3.1$~of~\cite{Ye2004Nondifferentiable}~and~\cite{Wu2016A}, we can use subgradient to deal with~$S\left(\bmu^{(m^*)},\btheta^{(m^*)}\right)$~and~$S^{(m^{*}+1)}\left(\bmu^{(m^*)},\btheta^{(m^*)}\right)$~as both of them are Lipschitz functions.

Since the subgradient of the~$S\left(\bmu^{(m^*)},\btheta^{(m^*)}\right)$~is the same as~$S^{(m^{*}+1)}\left(\bmu^{(m^*)},\btheta^{(m^*)}\right)$~and then we simply verify the KKT condition without inequality constraints, that is
\begin{align}\nonumber
&2(\bz^{*})^{\top}(\by^*-\bz^*\bmu)+\text{sgn}(\bmu)=\bm{0},\\\nonumber
&\lambda_2b_{ij }\frac{\btheta_{ij}}{||\btheta_{ij}||_2}+\rho(\btheta_{ij}-\bmu_i+\bmu_j+\bv_{ij})=0, \\\nonumber
&\btheta_{ij}=\bmu_{i}-\bmu_{j},
\end{align}
where~$\text{sgn}(\bmu)=(\text{sgn}(\mu_1),\text{sgn}(\mu_2),\cdots,\text{sgn}(\mu_p))^{\top}$, $\text{sgn}(x)$ is the sign function, $\bz=(\bz_1^{\top},\bz_2^{\top},\cdots,\bz_n^{\top})^{\top}$ and $b_{ij}=I(||\btheta_{ij}^{(m)}||_2<\tau)$. Note that~$\left(\hat{\bmu}^{(m)},\hat{\btheta}^{(m)}\right)$~is the limiting point of the ADMM iterations, thus the first equation holds under the cyclic coordinate descent algorithm since~the target function in (\ref{equaiton:2.5})~is convex. Similarly, the second equation holds under the soft thresholding operator. The last equation holds due to the fact that the steps of updating~$\bv_{ij}$~is the same as the KKT conditions. Thus $\left(\bmu^{(m^*)},\btheta^{(m^*)}\right)$~is a KKT point of~$S(\bmu,\btheta)$.

\bibliographystyle{model2-names}
\bibliography{ref}
	
\end{document}